\documentclass[conference]{IEEEtran}
\usepackage[T1]{fontenc}
\usepackage[utf8]{inputenc}
\usepackage{amssymb}
\usepackage[pdftex]{graphicx}
\usepackage{subcaption} 
\usepackage{amsmath}
\usepackage{mathtools}
\usepackage{array}
\usepackage[noadjust]{cite}
\usepackage{glossaries}
\usepackage{algorithm}
\usepackage{algpseudocode}
\usepackage{xcolor}
\usepackage{tikz}
\usepackage{pgfplots}

\pgfdeclarelayer{background}
\pgfdeclarelayer{foreground}
\pgfsetlayers{background,main,foreground}
\usepgfplotslibrary{fillbetween}
\usepgfplotslibrary{external}
\usetikzlibrary{spy,backgrounds}

\makeatletter
\let\MYcaption\@makecaption
\makeatother

\usepackage{subcaption}

\makeatletter
\let\@makecaption\MYcaption
\makeatother
\algnewcommand{\LineComment}[1]{\(\triangleright\) #1}

\newacronym{AWGN}{AWGN}{additive white Gaussian noise}
\newacronym{BER}{BER}{bit error rate}
\newacronym{BLER}{BLER}{block error rate}
\newacronym{BP}{BP}{backpropagation}
\newacronym{CE}{CE}{cross-entropy}
\newacronym{CFO}{CFO}{carrier frequency offset}
\newacronym{CP}{CP}{cyclic prefix}
\newacronym{CSI}{CSI}{channel state information}
\newacronym{DL}{DL}{deep learning}
\newacronym{DFT}{DFT}{discrete Fourier transform}
\newacronym{FFT}{FFT}{fast Fourier transform}
\newacronym{IFFT}{IFFT}{inverse fast Fourier transform}
\newacronym{GAN}{GAN}{generative adversarial network}
\newacronym{GPU}{GPU}{Graphics Processing Unit}
\newacronym{iid}{i.i.d.\@}{independent and identically distributed}
\newacronym{ISI}{ISI}{inter-symbol interference}
\newacronym{LOS}{LOS}{line-of-sight}
\newacronym{MDP}{MDP}{Markov decision process}
\newacronym{MIMO}{MIMO}{multiple-input multiple-output}
\newacronym{ML}{ML}{machine learning}
\newacronym{NN}{NN}{neural network}
\newacronym{OFDM}{OFDM}{orthogonal frequency-division multiplexing}
\newacronym{pdf}{pdf}{probability density function}
\newacronym{pmf}{pmf}{probability mass function}
\newacronym{RBF}{RBF}{Rayleigh block-fading}
\newacronym{ReLU}{ReLU}{rectified linear unit}
\newacronym{RL}{RL}{reinforcement learning}
\newacronym{SDR}{SDR}{software defined radio}
\newacronym{SER}{SER}{symbol error rate}
\newacronym{SNR}{SNR}{signal-to-noise ratio}
\newacronym{SINR}{SINR}{signal-to-interference-plus-noise ratio}
\newacronym{SGD}{SGD}{stochastic gradient descent}
\newacronym{wrt}{w.r.t.\@}{with respect to}
\newacronym{QPSK}{QPSK}{quaternary phase-shift keying}
\newacronym{SPSA}{SPSA}{simultaneous perturbation stochastic approximation}
\newacronym{SWIPT}{SWIPT}{simultaneous wireless information and power transfer}
\newacronym{MSE}{MSE}{mean squared error}
\newacronym{PSNR}{PSNR}{peak signal-to-noise ratio}
\newacronym{luib}{LUIB}{locally uniformly integrably bounded}
\newacronym{KL}{KL}{Kullback–Leibler}
\newacronym{MINE}{MINE}{mutual information neural estimator}
\newacronym{QAM}{QAM}{quadrature amplitude modulation}
\newacronym{PSK}{PSK}{phase-shift keying}
\newacronym{ASK}{ASK}{amplitude-shift keying}
\newacronym{LMMSE}{LMMSE}{linear minimum mean square error}
\renewcommand{\vec}[1]{\mathbf{#1}}
\newcommand{\vecs}[1]{\boldsymbol{#1}}

\newcommand{\bv}{\vec{b}}
\newcommand{\cv}{\vec{c}}

\newcommand{\sv}{\vec{s}}

\newcommand{\thetav}{\vecs{\theta}}

\newcommand{\Cm}{\vec{C}}

\newcommand{\Lc}{{\cal L}}

\newcommand{\Sc}{{\cal S}}

\newcommand{\CC}{\mathbb{C}}

\newcommand{\RR}{\mathbb{R}}

\newcommand{\tp}{^{\mathsf{T}}}

\newcommand{\LB}{\left(}
\newcommand{\RB}{\right)}
\newcommand{\LP}{\left\{}
\newcommand{\RP}{\right\}}

\renewcommand{\log}[1]{\mathop{\mathrm{log}}\LB #1\RB}
\renewcommand{\exp}[1]{\mathop{\mathrm{exp}}\LB #1\RB}

\newcommand{\EE}{{\mathbb{E}}}

 \newcommand{\argmax}[1]{\underset{#1}{\operatorname{arg}\,\operatorname{max}}\;}

\newcommand\abs[1]{\left| #1 \right|}

\begin{document}
\title{Joint Learning of Geometric and Probabilistic Constellation Shaping}


\IEEEoverridecommandlockouts 

\author{\IEEEauthorblockN{Maximilian Stark\IEEEauthorrefmark{1}\IEEEauthorrefmark{4}\IEEEauthorrefmark{5},
 Fay\c{c}al Ait Aoudia\IEEEauthorrefmark{3}\IEEEauthorrefmark{4}, and
Jakob Hoydis\IEEEauthorrefmark{3}
\thanks{\IEEEauthorrefmark{4}Equally contributed.}
\thanks{\IEEEauthorrefmark{5}Work carried out at Nokia Bell Labs France.}
}
\IEEEauthorblockA{\IEEEauthorrefmark{1}Hamburg University of Technology
Institute of Communications, Hamburg, Germany, Email: maximilian.stark@tuhh.de}
\IEEEauthorblockA{\IEEEauthorrefmark{3}Nokia Bell Labs, Paris, France, 
Email: \{faycal.ait\_aoudia, jakob.hoydis\}@nokia-bell-labs.com}}

\maketitle

\begin{abstract}
The choice of constellations largely affects the performance of communication systems.
When designing constellations, both the locations and probability of occurrence of the points can be optimized. These approaches are referred to as geometric and probabilistic shaping, respectively.
Usually, the geometry of the constellation is fixed, e.g., \gls{QAM} is used.
In such cases, the achievable information rate can still be improved by probabilistic shaping.
In this work, we show how autoencoders can be leveraged to perform probabilistic shaping of constellations.
We devise an information-theoretical description of autoencoders, which allows learning of capacity-achieving symbol distributions and constellations.
Recently, machine learning techniques to perform geometric shaping were proposed.
However, 
probabilistic shaping is more challenging as it requires the optimization of discrete distributions.
Furthermore, the proposed method enables joint probabilistic and geometric shaping of constellations over any channel model.
Simulation results show that the learned constellations achieve information rates very close to capacity on an \gls{AWGN} channel and outperform existing approaches on both \gls{AWGN} and fading channels.

\begin{IEEEkeywords}
Probabilistic shaping, Geometric shaping, Autoencoders
\end{IEEEkeywords}

\end{abstract}

\glsresetall

\section{Introduction}

Various constellation schemes were developed in  digital communications, including \gls{QAM}, \gls{PSK}, \gls{ASK} etc.
Shaping of constellations involves either optimizing the locations of the constellation points in the complex plane, i.e., geometric shaping, or optimizing the probabilities of occurrence of the constellation points, i.e., probabilistic shaping. In either case, the focal aim is to maximize the mutual information $I(X;Y)$ of the channel input $X$ and output $Y$ by optimizing the constellation.
This approach follows directly from the definition of the channel capacity $C$:
\begin{equation}
	\label{eq:cap}
	C = \underset{p(x)}{\max}~I(X;Y)
\end{equation}
where $p(x)$ denotes the marginal distribution of $X$.
Usually, finding the optimal $p(x)$ is a difficult problem as it requires the knowledge of the channel distribution $p(y|x)$.
Moreover, even if $p(y|x)$ is known, solving (\ref{eq:cap}) is often intractable.

In this work, we present how the recently proposed idea of end-to-end learning of communication systems by leveraging autoencoders~\cite{8054694} can be used to design constellations which maximize $I(X;Y)$, without requiring any tractable model of the channel.
Autoencoders have been used in the past to perform geometric shaping~\cite{8054694,jones2018geometric}.
However, to the best of our knowledge, leveraging autoencoders to achieve probabilistic shaping has not been explored.
In this paper, probabilistic shaping is learned by leveraging the recently proposed Gumbel-Softmax trick~\cite{jang2016categorical} to optimize discrete distributions.
Afterwards, joint geometric and probabilistic shaping of constellation is performed.
Presented results show that the achieved mutual information outperforms state-of-the-art systems over a wide range of \glspl{SNR} and on both \gls{AWGN} and fading channels, whereas current approaches are typically optimized to perform well only on specific channel models and small \gls{SNR} ranges~\cite{7307154}.

The rest of this paper is organized as follows.
Section~\ref{sec:ae} provides background on autoencoder-based communication systems and motivates their use for constellation shaping.
Section~\ref{sec:nn} details the considered \gls{NN} architecture and how the Gumbel-Softmax trick is leveraged to achieve probabilistic shaping.
Section~\ref{sec:res} provides results on the mutual information achieved by different schemes.
Finally, Section~\ref{sec:conclu} concludes this paper.

\subsection*{Notations}
Random variables are denoted by capital italic font, e.g., $X ,Y$ with realizations $x \in \mathcal{X},y \in \mathcal{Y}$, respectively. 
$I(X;Y)$, $p(y|x)$ and $p(x,y)$ represent the mutual information, conditional probability and joint probability distribution of the two random variables $X$ and $Y$. 
Vectors are represented using a lower case bold font, e.g., $\mathbf{y}$, upper case bold font letters denote matrices, e.g., $\mathbf{C}$.




\section{Autoencoder-based communication systems}
\label{sec:ae}

\begin{figure}
 	\centering
	\includegraphics[scale=0.30]{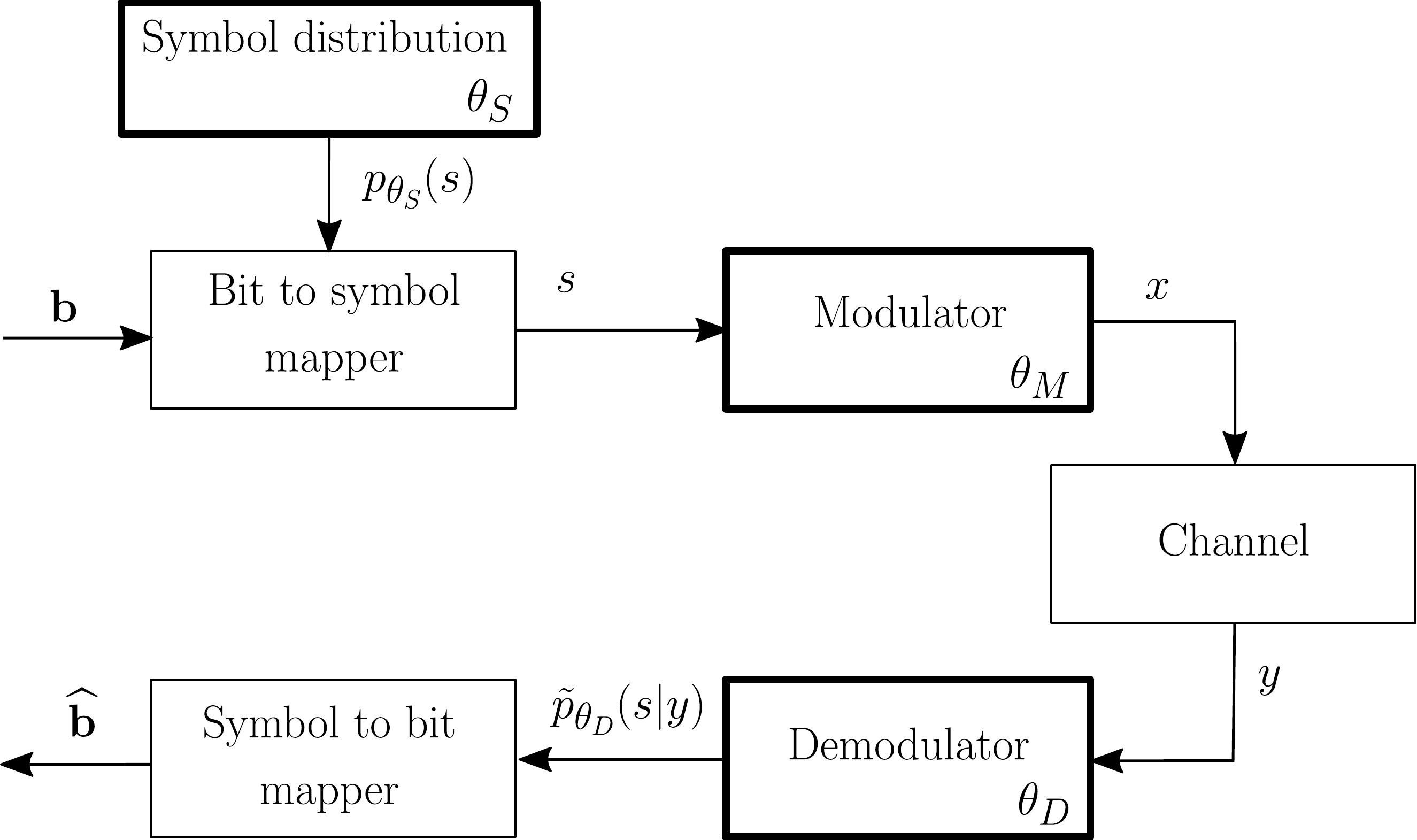}
	\caption{Trainable end-to-end communication system. Components on which this work focuses are indicated by thicker outlines.\label{fig:ae}}
\end{figure}

The key idea of autoencoder-based communication systems is to regard transmitter, channel, and receiver as a single \gls{NN} such that the transmitter and receiver can be optimized in an end-to-end manner.
This idea was pioneered in~\cite{8054694}, and has led to many extensions~\cite{kim2018communication,kimOFDM2018,osheamimo2017,bourtsoulatze2018deep}.
Fig.~\ref{fig:ae} shows the end-to-end communication system considered in this work.
The system takes as input a bit sequence denoted by $\bv$ which is mapped onto hypersymbols $s \in \mathcal{S}$ such that symbols $s$ appear with frequencies corresponding to a parametric distribution $p_{\thetav_S}(s)$ with parameters $\thetav_S$.
Here, $\Sc = \{1,\dots,N\}$ is the eventspace of the random variable $S$, $N$ being the modulation order.
The sequence of hypersymbols is fed into a symbol modulator which maps each symbol $s$ into a constellation point $x \in \CC$.
The modulator is implemented as an \gls{NN} $f_{\thetav_M}$ with trainable parameters $\thetav_M$.

The demodulator is also implemented as an \gls{NN} with trainable parameters $\thetav_D$, which maps each received sample $y \in \CC$ to a probability vector over the set of symbols $\Sc$.
The mapping defined by the demodulator is denoted by $\tilde{p}_{\thetav_D}(s|y)$, and defines, as it will be seen below, an approximation of the true posterior distribution $p_{\thetav_S,\thetav_M}(s|y)$.
Finally, the sent bits are reconstructed by the symbols to bits mapper from $\tilde{p}_{\thetav_D}(s|y)$.

\subsection{Mutual Information Perpective on Autoencoders }

In this work, it is assumed that a bits to symbols mapper exists, which maps the bits from $\bv$ to symbols $s \in \Sc$ according to the distribution $p_{\thetav_S}(s)$.
This can be done, e.g., using the algorithm presented in \cite{7322261}.
Therefore, in the rest of this work, the transmitter directly outputs the transmit symbols sampled from $p_{\thetav_S}(s)$, and the receiver aims to reconstruct the transmitted symbols by approximating the posterior distribution $p_{\thetav_S,\thetav_M}(s|y)$.
Thus, only the signal processing blocks surrounded by thicker outlines in Fig.~\ref{fig:ae} are of interest in this work.
The distribution of $X$ equals
\begin{equation}
p_{\thetav_S,\thetav_M}(x) = \sum_{s = 1}^N \delta\left(x - f_{\thetav_M}(s)\right)p_{\thetav_S}(s).
\end{equation}
where $\delta(.)$ denotes the Dirac distribution.
Please recall, that, as defined in \eqref{eq:cap}, the target of constellation shaping is to find $p_{\thetav_S}(s)$, such that $I(X;Y)$ is maximized.
One performs constellation shaping by optimizing $p_{\thetav_S}$ (probabilistic shaping) or $f_{\thetav_M}$ (geometric shaping) so that $I(X;Y)$ is maximized.

As the demodulator performs a classification task, for training, the categorical cross entropy 
\begin{align}
&\Lc(\thetav_S, \thetav_M,\thetav_D) \triangleq \EE_{s,y} \LP -\log{\tilde{p}_{\thetav_D}(s|y)} \RP\\
&= -\sum_{s=1}^N p_{\thetav_S}(s) \int_{y} p\left(y|f_{\thetav_M}(s)\right) \log{\tilde{p}_{\thetav_D}(s|y)} dy
\end{align}
is used as loss function.
Rewriting the loss function yields 
\begin{align} \label{eq:L}
\Lc(\thetav_S, \thetav_M,\thetav_D) &= 
H_{\thetav_S}(S) - I_{\thetav_S,\thetav_M}(X;Y) \nonumber\\
&+ \EE_{y} \LP \text{D}_{\text{KL}}\LB p_{\thetav_S,\thetav_M}(x|y)||\tilde{p}_{\thetav_D}(x|y) \RB \RP
\end{align}
where D\textsubscript{KL} is the \gls{KL} divergence. 
A more detailed derivation is given in the Appendix.

Notice that if only geometric shaping is performed, no optimization \gls{wrt} $\thetav_S$ is done and therefore the first term in~(\ref{eq:L}) is a constant.
However, when performing probabilistic shaping, minimizing $\Lc$ leads to the minimization of $H_{\thetav_S}(S)$.
To avoid this unwanted effect, we define the loss function
\begin{equation}
\widehat{\Lc}(\thetav_S,\thetav_M,\thetav_D) \triangleq \Lc(\thetav_S,\thetav_M,\thetav_D) - H_{\thetav_S}(S). \label{eq:lhat_exp}
\end{equation}
Training the end-to-end system by minimizing $\widehat{\Lc}$ corresponds to maximizing the mutual information of the channel inputs $X$ and outputs $Y$, while minimizing the \gls{KL} divergence between the true posterior distribution $p_{\thetav_S,\thetav_M}(x|y)$ and the one learned by the receiver $\tilde{p}_{\thetav_D}(x|y)$.
Moreover, the \gls{NN} implementing the receiver should approximate the posterior distribution $p_{\thetav_S,\thetav_M}(x|y)$ of a constellation maximizing the mutual information with high precision. This avoids learning a constellation where the posterior distribution is well approximated, but which does not maximize the mutual information.
In practice, this is ensured by choosing the \gls{NN} implementing the demodulator complex enough to ensure that the trainable receiver is capable of approximating a wide range of posterior distribution with high precision.

\begin{figure*}
	\centering
	\includegraphics[scale=0.35]{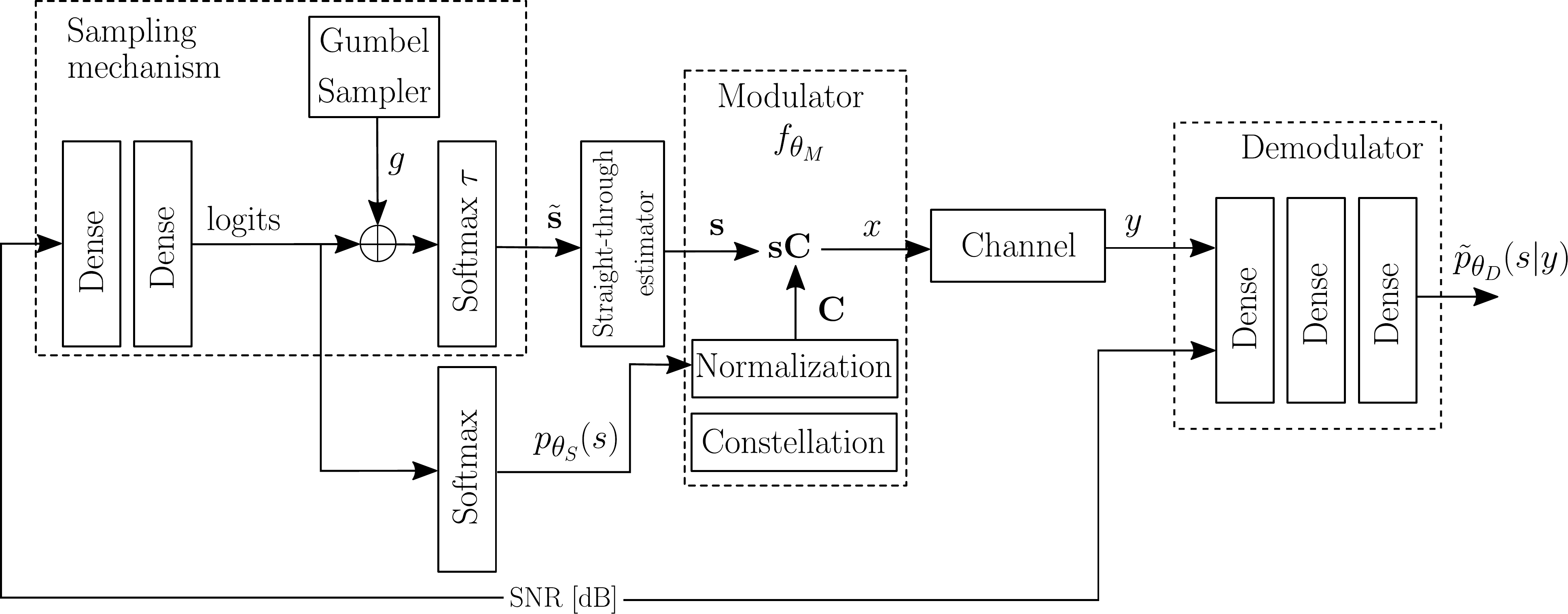}
	\caption{End-to-end system architecture\label{fig:ae_det}}
\end{figure*}

Recently, \cite{fritschek2019deep} proposed to leverage the \gls{MINE} approach as described in~\cite{belghazi2018mine} to train the transmitter such that the estimated mutual information of the input and output of the channel is maximized.
Therefore, training a transmitter with this approach is similar to training a transmitter as a part of an autoencoder, as in both cases the transmitter is trained to maximize the mutual information.
Optimization using \gls{MINE} does not require to train the receiver.
However, the autoencoder approach 
jointly learns the transmitter and the receiver including the corresponding posterior distribution. The respective soft information output by the learned receiver can then be used in subsequent units, e.g., a channel decoder. 
Moreover, whereas \gls{MINE} requires an additional \gls{NN} only to approximate the mutual information, using an autoencoder the mutual information can be estimated from the loss as $-\widehat{\Lc}$ provides a tight lower bound, assuming a sufficiently complex \gls{NN} implementing the receiver. 

As by training an autoencoder-based communication system one maximizes the mutual information of $X$ and $Y$, it can be used to perform constellation shaping.
Although, geometric shaping using autoencoders has been done in the past~\cite{8054694,jones2018geometric,8792076}, performing probabilistic shaping is less straightforward as it requires to optimize the sampling mechanism for symbols $s$ drawn from $\Sc$.

\section{Learning Constellation Shaping}
\label{sec:nn}

The end-to-end system considered in this work is presented in Fig.~\ref{fig:ae}.
This section details the architecture of each trainable element, i.e., the symbol distribution, the modulator and the demodulator.
Fig.~\ref{fig:ae_det} shows in detail the architecture of the considered end-to-end system.

\subsection{Symbols Distribution}
\label{sec:symbols}
The challenge of performing probabilistic shaping with machine learning-based algorithms comes from the difficulty of training a sampling mechanism for symbols $s$ drawn from the finite set $\Sc$.
This issue is addressed in this work by leveraging the Gumbel-Softmax trick~\cite{jang2016categorical}, an extension of the Gumbel-Max trick~\cite{hazan2012partition}.
The Gumbel-Max trick provides a convenient way to sample a discrete distribution $p_{\thetav_S}(s)$, by computing the samples as follows:
\begin{equation}
	s = \argmax{i=1,\dots,S} \LB g_i + \log{p_{\thetav_S}(i)} \RB
\end{equation}
where $g_i$ are \gls{iid} samples drawn from a standard Gumbel distribution.
Because the $\operatorname{arg~max}$ operator is not differentiable, one cannot train $p_{\thetav_S}(s)$ using usual \gls{SGD} methods.
The key idea of the Gumbel-Softmax trick is to use the softmax function as a differentiable approximation to $\operatorname{arg~max}$.
More precisely, one generates a vector of dimension $N$, denoted by $\tilde{\sv}$, with components
\begin{equation}
	\label{eq:gumble_softmax}
	\tilde{s}_i = \frac{\exp{\left(g_i + \log{p_{\thetav_S}(i)}\right)/\tau}}{\sum_{j=1}^S \exp{\left(g_j + \log{p_{\thetav_S}(j)}\right)/\tau}},~i=1,\dots,N
\end{equation}
where $\tau$ is a positive parameter called the \emph{temperature}.
$\tilde{\sv}$ is a probability vector which is such that $\argmax{i}~\tilde{s_i} = s$.
It is an approximation of the \emph{one-hot} representation of $s$ denoted by $\sv$, i.e., the $N-$dimensional vector for which all elements are set to zero except the $s$th which is set to one.
As the temperature goes to zero, samples generated by the Gumbel-Softmax method become closer to one-hot vectors, and their distribution becomes closer to $p_{\thetav_S}(s)$~\cite{jang2016categorical}.

Fig.~\ref{fig:ae_det} shows the architecture of the sampling mechanism.
The optimal probabilistic shaping depends on the \gls{SNR} of the respective channel, which therefore must be \emph{a priori} known by the transmitter~\cite{7307154}.
The \gls{SNR} (in dB) is fed to an \gls{NN} with trainable parameters $\thetav_S$, and therefore the \gls{NN} generates a continuum of distributions $p_{\thetav_S}$ that are determined by the \gls{SNR}.
The \gls{NN} is made of two dense layers, and generates the \emph{logits} of the symbols distribution $p_{\thetav_S}(s)$.
The logits are the unormalized log probabilities, and the distribution $p_{\thetav_S}(s)$ can be retrieved by applying a softmax activation to the logits.
The first dense layer is made of 128 units with ReLU activations, and the second layer of $N$ units with linear activations.
By tuning the \gls{NN} parameters $\thetav_S$, one therefore optimizes the distribution $p_{\thetav_S}(s)$.
The Gumbel-Softmax trick is then applied to the logits.

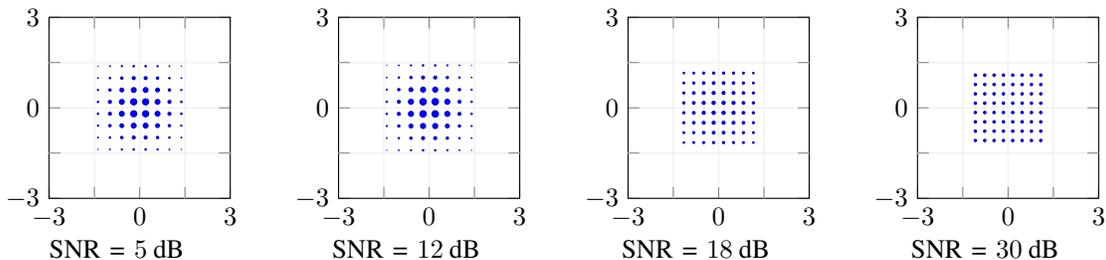
\begin{figure*}
\centering
\begin{tabular}{c c c c}
	\raisebox{-.5\height}{
       \begin{tikzpicture}
            \begin{axis}[%
            name=plot21,
            grid=both,
            grid style={line width=.6pt, draw=gray!10},
            only marks,
            width=0.22\textwidth,
            height=0.22\textwidth,
            xmin=-3, xmax=3,
            ymin=-3, ymax=3,
            xtick={-3,0,3},
            ytick={-3,0,3},
            xticklabels={$-3$, $0$, $3$},
            yticklabels={$-3$, $0$, $3$},
            extra x ticks={-1.5, 1.5},
            extra y ticks={-1.5, 1.5},
            extra x tick labels={},
            extra y tick labels={}
            ]
                \addplot[
                    scatter,
                    only marks,
                    scatter src=explicit symbolic,
		            visualization depends on={\thisrow{s} \as \perpointmarksize},
		            scatter/@pre marker code/.append style={
		                /tikz/mark size={\perpointmarksize*30}
		            }
                ] table [x=x, y=y, col sep=comma, meta=s] {figs/psqam_m64_const_SNR5.0.csv};
            \end{axis}
        \end{tikzpicture}}&
        \raisebox{-.5\height}{
        \begin{tikzpicture}
            \begin{axis}[%
            name=plot22,
            grid=both,
            grid style={line width=.6pt, draw=gray!10},
            only marks,
            width=0.22\textwidth,
            height=0.22\textwidth,
            xmin=-3, xmax=3,
            ymin=-3, ymax=3,
            xtick={-3,0,3},
            ytick={-3,0,3},
            xticklabels={$-3$, $0$, $3$},
            yticklabels={$-3$, $0$, $3$},
            extra x ticks={-1.5, 1.5},
            extra y ticks={-1.5, 1.5},
            extra x tick labels={},
            extra y tick labels={}
            ]
                \addplot[
                    scatter,
                    only marks,
                    scatter src=explicit symbolic,
		            visualization depends on={\thisrow{s} \as \perpointmarksize},
		            scatter/@pre marker code/.append style={
		                /tikz/mark size={\perpointmarksize*30}
		            }
                ] table [x=x, y=y, col sep=comma, meta=s] {figs/psqam_m64_const_SNR15.0.csv};
            \end{axis}
        \end{tikzpicture}}&
        \raisebox{-.5\height}{
        \begin{tikzpicture}
            \begin{axis}[%
            name=plot23,
            grid=both,
            grid style={line width=.6pt, draw=gray!10},
            only marks,
            width=0.22\textwidth,
            height=0.22\textwidth,
            xmin=-3, xmax=3,
            ymin=-3, ymax=3,
            xtick={-3,0,3},
            ytick={-3,0,3},
            xticklabels={$-3$, $0$, $3$},
            yticklabels={$-3$, $0$, $3$},
            extra x ticks={-1.5, 1.5},
            extra y ticks={-1.5, 1.5},
            extra x tick labels={},
            extra y tick labels={}
            ]
                \addplot[
                    scatter,
                    only marks,
                    scatter src=explicit symbolic,
		            visualization depends on={\thisrow{s} \as \perpointmarksize},
		            scatter/@pre marker code/.append style={
		                /tikz/mark size={\perpointmarksize*30}
		            }
                ] table [x=x, y=y, col sep=comma, meta=s] {figs/psqam_m64_const_SNR20.0.csv};
            \end{axis}
        \end{tikzpicture}}&
        \raisebox{-.5\height}{
                \begin{tikzpicture}
            \begin{axis}[%
            name=plot24,
            grid=both,
            grid style={line width=.6pt, draw=gray!10},
            only marks,
            width=0.22\textwidth,
            height=0.22\textwidth,
            xmin=-3, xmax=3,
            ymin=-3, ymax=3,
            xtick={-3,0,3},
            ytick={-3,0,3},
            xticklabels={$-3$, $0$, $3$},
            yticklabels={$-3$, $0$, $3$},
            extra x ticks={-1.5, 1.5},
            extra y ticks={-1.5, 1.5},
            extra x tick labels={},
            extra y tick labels={}
            ]
                \addplot[
                    scatter,
                    only marks,
                    scatter src=explicit symbolic,
		            visualization depends on={\thisrow{s} \as \perpointmarksize},
		            scatter/@pre marker code/.append style={
		                /tikz/mark size={\perpointmarksize*30}
		            }
                ] table [x=x, y=y, col sep=comma, meta=s] {figs/psqam_m64_const_SNR25.0.csv};
            \end{axis}
        \end{tikzpicture}}\\
    SNR = $5\:$dB & SNR = $12\:$dB & SNR = $18\:$dB & SNR = $30\:$dB\\
\end{tabular}
\caption{Learned probabilistic shaping for $N = 64$. The size of the points is proportional to their probabilities of occurrence.\label{fig:const_ps}}
\end{figure*}

\subsection{Modulator}
\label{sec:modulator}

The modulator is made of a matrix of dimension $N \times 2$ followed by a normalization layer, as shown in Fig.~\ref{fig:ae_det}.
The matrix consists of the unnormalized constellation point locations.
The normalized constellation is denoted by $\Cm = [\cv_1,\dots,\cv_i,\dots,\cv_{N}]\tp$ where $\cv_i \in \RR^2,~i=1,\dots,N$.
By taking the product of a one-hot vector $\sv$ with the $s$th element set to one and $\Cm$, one selects a constellation point.
Normalization is performed to ensure the expected energy of the constellation equals unity, i.e.,
\begin{equation}
	\sum_{s \in \mathcal{S}} p_{\thetav_S}(s) \abs{x_s}^2_2 = 1.
\end{equation}
If only probabilistic shaping is performed, the constellation is not trained and some fixed constellation, e.g., \gls{QAM}, is used.
When geometric shaping is performed, the constellation is trainable, and $\thetav_M$ corresponds to the unnormalized constellation points.

A drawback of the Gumbel-Softmax trick is that the generated vector $\tilde{\sv}$ is only an approximation of a true one-hot vector $\sv$.
As a consequence, taking the product of $\tilde{\sv}$ and $\Cm$ results in a convex combination of multiple constellation points $\cv_s$.
To avoid this issue, we take advantage of the straight-through estimator~\cite{bengio2013estimating}, which uses the true one-hot vectors $\sv$ for the forward pass and the approximate one-hot vector $\tilde{\sv}$ for the backward pass at training.

\subsection{Demodulator}
\label{sec:demodulator}
The trainable demodulator consists of three dense layers, as shown in Fig.~\ref{fig:ae_det}.
The first two layers are made of 128 units with ReLU activations, while the last layer is made of $N$ units with softmax activation, to output a probability vector over the set of symbols $\Sc$.
As opposed to prior art, the demodulator takes as input the \gls{SNR} (in dB).
This was motivated by the observation that the posterior distribution depends on the \gls{SNR}, and it was experimentally found out to be crucial to achieve the best performance.

\section{Simulations Results}
\label{sec:res}

In this section, the mutual information of the channel input and output achieved by the proposed scheme is compared to state-of-the-art modulation schemes considering \gls{AWGN} and Rayleigh channels.
Training of the end-to-end system introduced in the previous section is performed \gls{wrt} to the loss function $\widehat{\Lc}$ defined in~(\ref{eq:lhat_exp}), as opposed to previous works which train autoencoders \gls{wrt} to the usual cross-entropy $\Lc$ defined in~(\ref{eq:L}).
Using $\widehat{\Lc}$ as loss function is crucial to enable probabilistic shaping, as using $\Lc$ would rather lead to a minimization of the source entropy $H_{\thetav_S}(S)$ instead of maximization of the mutual information $I(X;Y)$.
The autoencoder was implemented with the TensorFlow framework~\cite{tensorflow2015-whitepaper}.
Training was performed with the Adam \gls{SGD} variant~\cite{Kingma15}, with batch sizes progressively increasing from 100 to 10000, and learning rates progressively decreasing from $10^{-3}$ to $10^{-5}$.
When probabilistic shaping was performed, the temperature in \eqref{eq:gumble_softmax} was set to 10.
First, we present the results for learned probabilistic shaping of a QAM modulation.
Afterwards, we show the results obtained when both the locations and probabilities of the constellation points are optimized, i.e., joint geometric and probabilistic shaping.
The considered modulation orders $N$ are 16, 64, 256, and 1024.

\subsection{Probabilistic Shaping over the \gls{AWGN} Channel}

Probabilistic shaping of \gls{QAM} is studied in great detail in~\cite{7307154}.
It is well know~\cite{kschischang1993optimal} that for an \gls{AWGN} channel, distributions $p(s)$ from the Maxwell-Boltzmann family maximize the mutual information $I(X;Y)$ and, thus, allow to achieve the capacity in a certain regime.
However, in this work, we do not enforce the learning of distributions from this family.
The following modulation schemes are compared: \gls{QAM} with no probabilistic shaping, \gls{QAM} with probabilistic shaping optimized as proposed in this work, and \gls{QAM} with Maxwell-Boltzmann distribution as probabilistic shaping as in~\cite{7307154}.
\gls{QAM} with Maxwell-Boltzmann from~\cite{7307154} is only presented for modulation orders of 16, 64, and 256, and the distributions are optimized for specific \gls{SNR} values.
Notice that the proposed approach has the benefit of training an \gls{NN} which computes optimized shaping distributions over a wide range of \glspl{SNR}.
For this evaluation, the sampling mechanism was trained for \gls{SNR} values ranging from $-2\:$dB to $40\:$dB.

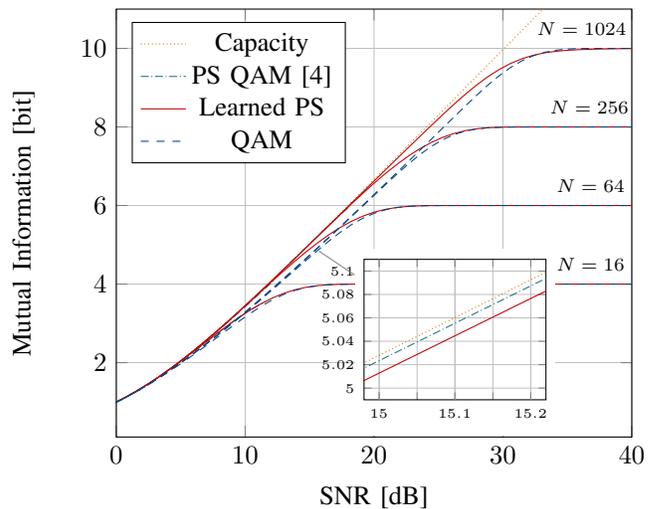
\begin{figure}

		\begin{tikzpicture}[every pin/.style={fill=white}]
		\definecolor{DarkBlue}{rgb}{0, 0.2706, 0.541}
		\definecolor{DarkOrange}{rgb}{0.92, 0.43, 0}
		\definecolor{Green}{rgb}{0.098, 0.4784, 0.5176}
		\definecolor{DarkRed}{rgb}{0.7529, 0, 0}
		
			\begin{axis}[
				grid=both,
				grid style={line width=.4pt, draw=gray!10},
				major grid style={line width=.2pt,draw=gray!50},
				xlabel={SNR [dB]},
				ylabel={Mutual Information [bit]},
				ymax=11,
				xmax=40,
				xmin=0,
				legend pos = north west,
			]

				\addplot[thin,DarkOrange,densely dotted] table [x=snr, y=cap, col sep=comma] {figs/m16.csv};

				\addplot[thin,densely dash dot,Green] table [x=snr, y=paper, col sep=comma] {figs/QAM16_AWGN_shaping_paper.csv};
				\addplot[thin,densely dash dot,Green,forget plot] table [x=snr, y=paper, col sep=comma] {figs/QAM64_AWGN_shaping_paper.csv};
				\addplot[thin,densely dash dot,Green,forget plot] table [x=snr, y=paper, col sep=comma] {figs/QAM256_AWGN_shaping_paper.csv};
			
				\addplot[thin,DarkRed] table [x=snr, y=ps, col sep=comma] {figs/m16.csv};
				\addplot[thin,DarkRed,forget plot] table [x=snr, y=ps, col sep=comma] {figs/m64.csv};
				\addplot[thin,DarkRed,forget plot] table [x=snr, y=ps, col sep=comma] {figs/m256.csv};
				\addplot[thin,DarkRed,forget plot] table [x=snr, y=ps, col sep=comma] {figs/m1024.csv};				
				
				\addplot[thin,DarkBlue,dashed] table [x=snr, y=qam, col sep=comma] {figs/m16.csv};				
				\addplot[thin,DarkBlue,densely dashed,forget plot] table [x=snr, y=qam, col sep=comma] {figs/m64.csv};				
				\addplot[thin,DarkBlue,densely dashed,forget plot] table [x=snr, y=qam, col sep=comma] {figs/m256.csv};				
				\addplot[thin,DarkBlue,densely dashed,forget plot] table [x=snr, y=qam, col sep=comma] {figs/m1024.csv};

				\node[text width=2cm] at (axis cs:40,4.5) {\scriptsize $N = 16$};
				\node[text width=2cm] at (axis cs:40,6.5) {\scriptsize $N = 64$};
				\node[text width=2cm] at (axis cs:39.5,8.5) {\scriptsize $N = 256$};
				\node[text width=2cm] at (axis cs:38.5,10.5) {\scriptsize $N = 1024$};

				\addlegendentry{Capacity}
				\addlegendentry{PS QAM \cite{7307154}}
				\addlegendentry{Learned PS}
				\addlegendentry{QAM}

				\coordinate (pt) at (axis cs:15,5);

				\end{axis}

				\node[pin=-5:{%
				    \begin{tikzpicture}[baseline,trim axis left,trim axis right]
				    \begin{axis}[
				        tiny,
					grid=both,
					grid style={line width=.4pt, draw=gray!10},
					major grid style={line width=.2pt,draw=gray!50},
				      xmin=15,xmax=15.2,
				      ymin=5,ymax=5.1,
				      enlargelimits,
				      xticklabels={},
					  extra x ticks={15,15.1,15.2},
				    ]
				    \addplot[thin,DarkRed,forget plot] table [x=snr, y=ps, col sep=comma] {figs/m256.csv};	
				\addplot[thin,densely dash dot,Green,forget plot] table [x=snr, y=paper, col sep=comma] {figs/QAM256_AWGN_shaping_paper.csv};
				\addplot[thin,DarkOrange,densely dotted] table [x=snr, y=cap, col sep=comma] {figs/m16.csv};
				    \end{axis}
				    \end{tikzpicture}%
				}] at (pt) {};
							
			\end{tikzpicture}
	\caption{Mutual information achieved by the reference schemes \label{fig:mi_one} and the learned probabilistic shaping on the AWGN channel. Magnification is done for $N = 256$.}
\end{figure}

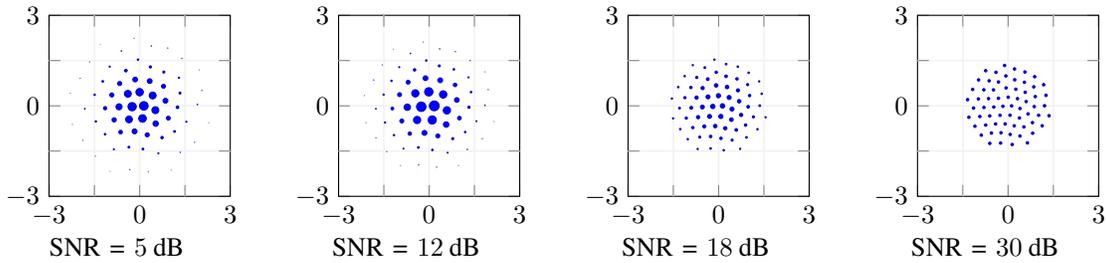
\begin{figure*}
\centering
\begin{tabular}{c c c c}
	\raisebox{-.5\height}{
       \begin{tikzpicture}
            \begin{axis}[%
            name=plot21,
            grid=both,
            grid style={line width=.6pt, draw=gray!10},
            only marks,
            width=0.22\textwidth,
            height=0.22\textwidth,
            xmin=-3, xmax=3,
            ymin=-3, ymax=3,
            xtick={-3,0,3},
            ytick={-3,0,3},
            xticklabels={$-3$, $0$, $3$},
            yticklabels={$-3$, $0$, $3$},
            extra x ticks={-1.5, 1.5},
            extra y ticks={-1.5, 1.5},
            extra x tick labels={},
            extra y tick labels={}
            ]
                \addplot[
                    scatter,
                    only marks,
                    scatter src=explicit symbolic,
		            visualization depends on={\thisrow{s} \as \perpointmarksize},
		            scatter/@pre marker code/.append style={
		                /tikz/mark size={\perpointmarksize*30}
		            }
                ] table [x=x, y=y, col sep=comma, meta=s] {figs/m64_const_SNR5.0.csv};
            \end{axis}
        \end{tikzpicture}}&
        \raisebox{-.5\height}{
        \begin{tikzpicture}
            \begin{axis}[%
            name=plot22,
            grid=both,
            grid style={line width=.6pt, draw=gray!10},
            only marks,
            width=0.22\textwidth,
            height=0.22\textwidth,
            xmin=-3, xmax=3,
            ymin=-3, ymax=3,
            xtick={-3,0,3},
            ytick={-3,0,3},
            xticklabels={$-3$, $0$, $3$},
            yticklabels={$-3$, $0$, $3$},
            extra x ticks={-1.5, 1.5},
            extra y ticks={-1.5, 1.5},
            extra x tick labels={},
            extra y tick labels={}
            ]
                \addplot[
                    scatter,
                    only marks,
                    scatter src=explicit symbolic,
		            visualization depends on={\thisrow{s} \as \perpointmarksize},
		            scatter/@pre marker code/.append style={
		                /tikz/mark size={\perpointmarksize*30}
		            }
                ] table [x=x, y=y, col sep=comma, meta=s] {figs/m64_const_SNR12.0.csv};
            \end{axis}
        \end{tikzpicture}}&
        \raisebox{-.5\height}{
        \begin{tikzpicture}
            \begin{axis}[%
            name=plot23,
            grid=both,
            grid style={line width=.6pt, draw=gray!10},
            only marks,
            width=0.22\textwidth,
            height=0.22\textwidth,
            xmin=-3, xmax=3,
            ymin=-3, ymax=3,
            xtick={-3,0,3},
            ytick={-3,0,3},
            xticklabels={$-3$, $0$, $3$},
            yticklabels={$-3$, $0$, $3$},
            extra x ticks={-1.5, 1.5},
            extra y ticks={-1.5, 1.5},
            extra x tick labels={},
            extra y tick labels={}
            ]
                \addplot[
                    scatter,
                    only marks,
                    scatter src=explicit symbolic,
		            visualization depends on={\thisrow{s} \as \perpointmarksize},
		            scatter/@pre marker code/.append style={
		                /tikz/mark size={\perpointmarksize*30}
		            }
                ] table [x=x, y=y, col sep=comma, meta=s] {figs/m64_const_SNR18.0.csv};
            \end{axis}
        \end{tikzpicture}}&
        \raisebox{-.5\height}{
                \begin{tikzpicture}
            \begin{axis}[%
            name=plot24,
            grid=both,
            grid style={line width=.6pt, draw=gray!10},
            only marks,
            width=0.22\textwidth,
            height=0.22\textwidth,
            xmin=-3, xmax=3,
            ymin=-3, ymax=3,
            xtick={-3,0,3},
            ytick={-3,0,3},
            xticklabels={$-3$, $0$, $3$},
            yticklabels={$-3$, $0$, $3$},
            extra x ticks={-1.5, 1.5},
            extra y ticks={-1.5, 1.5},
            extra x tick labels={},
            extra y tick labels={}
            ]
                \addplot[
                    scatter,
                    only marks,
                    scatter src=explicit symbolic,
		            visualization depends on={\thisrow{s} \as \perpointmarksize},
		            scatter/@pre marker code/.append style={
		                /tikz/mark size={\perpointmarksize*30}
		            }
                ] table [x=x, y=y, col sep=comma, meta=s] {figs/m64_const_SNR30.0.csv};
            \end{axis}
        \end{tikzpicture}}\\

    SNR = $5\:$dB & SNR = $12\:$dB & SNR = $18\:$dB & SNR = $30\:$dB\\
\end{tabular}
\caption{Learned joint shaping for $N = 64$. The size of the points is proportional to their probabilities of occurrence.\label{fig:const}}
\end{figure*}

The learned probability distributions are depicted for $N = 64$ in Fig.~\ref{fig:const_ps} and for several \glspl{SNR}.
The size of the points is proportional to their probabilities of occurrence.
It can be seen that the learned distributions look similar to a circular two-dimensional Gaussian distribution.
With increasing \glspl{SNR}, the learned distribution approaches a uniform distribution. 
In Fig.~\ref{fig:mi_one}, the mutual information $I(X;Y)$ achieved by the proposed scheme (cf. red solid curve) is compared to several reference schemes.
First \gls{QAM} with no probabilistic shaping is considered (cf. blue dashed curve).
Clearly, the proposed schemes outperform non-shaped \gls{QAM}.
Furthermore, Fig.~\ref{fig:mi_one} includes the mutual information curves of the probabilistic shaping scheme proposed in~\cite{7307154} (cf. green dash dotted curve).
Interestingly, the learned scheme is able to achieve a performance close to the probabilistic shaping from~\cite{7307154}, which leverages the optimum Maxwell-Boltzmann distribution.
Notice that no assumption on the channel was made to achieve this result.

\subsection{Joint Shaping over the \gls{AWGN} Channel}

\begin{figure}

		\begin{tikzpicture}[every pin/.style={fill=white}]
		\definecolor{DarkBlue}{rgb}{0, 0.2706, 0.541}
		\definecolor{DarkOrange}{rgb}{0.92, 0.43, 0}
		\definecolor{Green}{rgb}{0.098, 0.4784, 0.5176}
		\definecolor{DarkRed}{rgb}{0.7529, 0, 0}
		
			\begin{axis}[
				grid=both,
				grid style={line width=.4pt, draw=gray!10},
				major grid style={line width=.2pt,draw=gray!50},
				xlabel={SNR [dB]},
				ylabel={Mutual Information [bit]},
				ymax=11,
				xmax=40,
				xmin=0,
				legend pos = north west,
			]

				\addplot[thin,DarkOrange,densely dotted] table [x=snr, y=cap, col sep=comma] {figs/m16.csv};

				\addplot[thin,DarkRed] table [x=snr, y=psgs, col sep=comma] {figs/m16.csv};
				\addplot[thin,DarkRed,forget plot] table [x=snr, y=psgs, col sep=comma] {figs/m64.csv};
				\addplot[thin,DarkRed,forget plot] table [x=snr, y=psgs, col sep=comma] {figs/m256.csv};
				\addplot[thin,DarkRed,forget plot] table [x=snr, y=psgs, col sep=comma] {figs/m1024.csv};

				\addplot[semithick,densely dash dot,Green] table [x=snr, y=gs, col sep=comma] {figs/m16.csv};
				\addplot[thin,densely dash dot,Green,forget plot] table [x=snr, y=gs, col sep=comma] {figs/m64.csv};
				\addplot[thin,densely dash dot,Green,forget plot] table [x=snr, y=gs, col sep=comma] {figs/m256.csv};
				\addplot[thin,densely dash dot,Green,forget plot] table [x=snr, y=gs, col sep=comma] {figs/m1024.csv};
				
				\addplot[thin,DarkBlue,densely dashed] table [x=snr, y=qam, col sep=comma] {figs/m16.csv};
				\addplot[thin,DarkBlue,densely dashed,forget plot] table [x=snr, y=qam, col sep=comma] {figs/m64.csv};
				\addplot[thin,DarkBlue,densely dashed,forget plot] table [x=snr, y=qam, col sep=comma] {figs/m256.csv};
				\addplot[thin,DarkBlue,densely dashed,forget plot] table [x=snr, y=qam, col sep=comma] {figs/m1024.csv};

				\node[text width=2cm] at (axis cs:40,4.5) {\scriptsize $N = 16$};
				\node[text width=2cm] at (axis cs:40,6.5) {\scriptsize $N = 64$};
				\node[text width=2cm] at (axis cs:39.5,8.5) {\scriptsize $N = 256$};
				\node[text width=2cm] at (axis cs:38.5,10.5) {\scriptsize $N = 1024$};

				\addlegendentry{Capacity}
				\addlegendentry{Learned PS+GS}	
				\addlegendentry{GS}	
				\addlegendentry{QAM}
			
			\coordinate (pt) at (axis cs:15,5);
			
				\end{axis}

				\node[pin=-5:{%
				    \begin{tikzpicture}[baseline,trim axis left,trim axis right]
				    \begin{axis}[
				        tiny,
					grid=both,
					grid style={line width=.4pt, draw=gray!10},
					major grid style={line width=.2pt,draw=gray!50},
				      xmin=15,xmax=15.2,
				      ymin=5,ymax=5.1,
				      enlargelimits,
				      xticklabels={},
					  extra x ticks={15,15.1,15.2},
				    ]

				    \addplot[thin,DarkRed,forget plot] table [x=snr, y=psgs, col sep=comma] {figs/m256.csv};
				    \addplot[thin,densely dash dot,Green,forget plot] table [x=snr, y=gs, col sep=comma] {figs/m256.csv};
				    \addplot[thin,DarkOrange,densely dotted] table [x=snr, y=cap, col sep=comma] {figs/m16.csv};

				    \end{axis}
				    \end{tikzpicture}%
				}] at (pt) {};
							
			\end{tikzpicture}

			\caption{Mutual information achieved by the reference schemes \label{fig:mi_one_gs} and the learned joint probabilistic and geometric shaping on the AWGN channel. Magnification is done for $N = 256$.}

\end{figure}
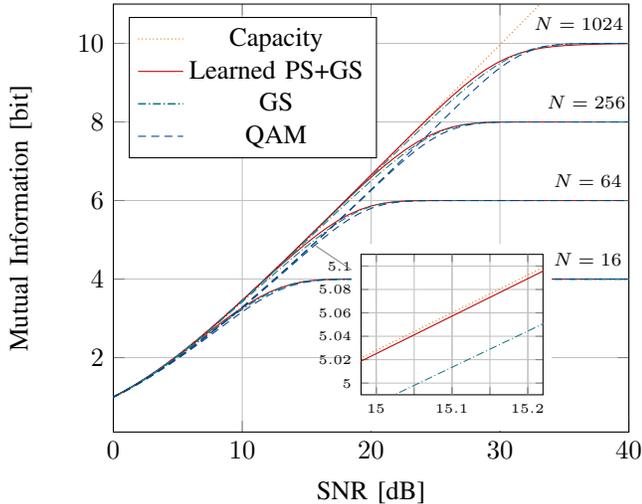

In this section, both the probability distribution and the geometry are assumed to be trainable.
Like in the previous section, the modulation orders 16, 64, 256, and 1024 and an AWGN channel are considered.
Fig.~\ref{fig:const} shows the joint probabilistic and geometric constellations for various \gls{SNR} values and for $M = 64$.
It can be observed that the learned shaping is similar to a two-dimensional Gaussian distribution.
For lower \glspl{SNR}, the learned shaping favors constellation points closer to the origin.
Due to the normalization, which ensures that $\EE\{\abs{x}^2\} = 1$, the less frequently transmitted outer points are placed further apart form the origin.
As the SNR increases, the distribution becomes uniform. 

In Fig.~\ref{fig:mi_one_gs}, the mutual information $I(X;Y)$ achieved by the proposed joint shaping scheme (cf. red solid curve) is compared to several reference schemes.
First, \gls{QAM} with no probabilistic shaping is considered (cf. blue dashed curve).
As for probabilistic shaping alone, the proposed joint shaping outperforms non-shaped \gls{QAM}.
Furthermore, Fig.~\ref{fig:mi_one} includes the mutual information curves performing only geometric shaping (cf. green dash dotted curve).
Joint geometric and probabilistic is also superior to geometric shaping alone.
Also, one observes that for all the modulation orders, probabilistic shaping achieves higher rates than geometric shaping, showing that probabilistic shaping of \gls{QAM} constellations enables higher gains than geometric shaping without probabilistic shaping.

It can be seen that the proposed \gls{NN} learns a joint probabilistic and geometric shaping which operates very close to capacity for a wide range of \glspl{SNR}.
From Fig.~\ref{fig:mi_one} and Fig.~\ref{fig:mi_one_gs}, one observes that the achievable gains enabled by constellation shaping are more significant the higher the modulation order gets.
In addition, comparing Fig.~\ref{fig:mi_one} and Fig.~\ref{fig:mi_one_gs} reveals that for all the modulation orders, the highest mutual information is achieved by the joint geometric and probabilistic scheme, which outperforms \mbox{PS-QAM} from~\cite{7307154}.

\subsection{Joint Shaping over the Rayleigh Channel}

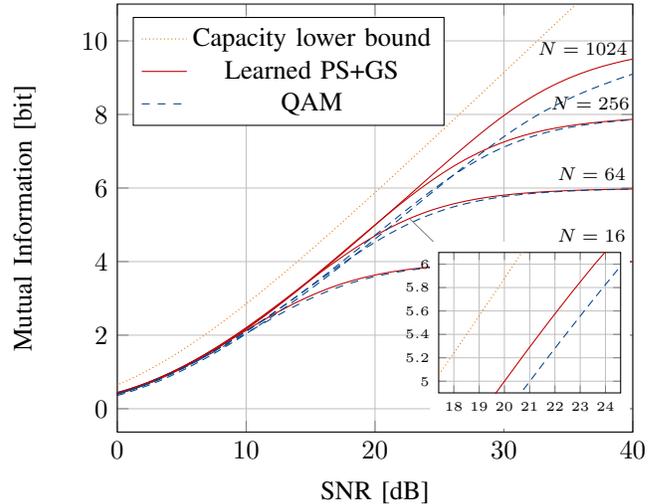
\begin{figure}

		\begin{tikzpicture}[every pin/.style={fill=white}]
		\definecolor{DarkBlue}{rgb}{0, 0.2706, 0.541}
		\definecolor{DarkOrange}{rgb}{0.92, 0.43, 0}
		\definecolor{Green}{rgb}{0.098, 0.4784, 0.5176}
		\definecolor{DarkRed}{rgb}{0.7529, 0, 0}
		
			\begin{axis}[
				grid=both,
				grid style={line width=.4pt, draw=gray!10},
				major grid style={line width=.2pt,draw=gray!50},
				xlabel={SNR [dB]},
				ylabel={Mutual Information [bit]},
				ymax=11,
				xmax=40,
				xmin=0,
				legend pos = north west,
			]

				\addplot[thin,DarkOrange,densely dotted] table [x=snr, y=cap, col sep=comma] {figs/m16_ray.csv};
			
				\addplot[thin,DarkRed] table [x=snr, y=psgs, col sep=comma] {figs/m16_ray.csv};
				\addplot[thin,DarkRed,forget plot] table [x=snr, y=psgs, col sep=comma] {figs/m64_ray.csv};
				\addplot[thin,DarkRed,forget plot] table [x=snr, y=psgs, col sep=comma] {figs/m256_ray.csv};
				\addplot[thin,DarkRed,forget plot] table [x=snr, y=psgs, col sep=comma] {figs/m1024_ray.csv};				
				
				\addplot[thin,DarkBlue,dashed] table [x=snr, y=qam, col sep=comma] {figs/m16_ray.csv};				
				\addplot[thin,DarkBlue,densely dashed,forget plot] table [x=snr, y=qam, col sep=comma] {figs/m64_ray.csv};				
				\addplot[thin,DarkBlue,densely dashed,forget plot] table [x=snr, y=qam, col sep=comma] {figs/m256_ray.csv};				
				\addplot[thin,DarkBlue,densely dashed,forget plot] table [x=snr, y=qam, col sep=comma] {figs/m1024_ray.csv};

				\node[text width=2cm] at (axis cs:40,4.7) {\scriptsize $N = 16$};
				\node[text width=2cm] at (axis cs:40,6.4) {\scriptsize $N = 64$};
				\node[text width=2cm] at (axis cs:39.5,8.3) {\scriptsize $N = 256$};
				\node[text width=2cm] at (axis cs:38.5,9.8) {\scriptsize $N = 1024$};

				\addlegendentry{Capacity lower bound}
				\addlegendentry{Learned PS+GS}
				\addlegendentry{QAM}

				\coordinate (pt) at (axis cs:22,5.4);

				\end{axis}

				\node[pin={[pin distance=0.4cm]-48:{%
				    \begin{tikzpicture}[baseline,trim axis left,trim axis right]
				    \begin{axis}[
				        tiny,
					grid=both,
					grid style={line width=.4pt, draw=gray!10},
					major grid style={line width=.2pt,draw=gray!50},
				      xmin=18,xmax=24,
				      ymin=5,ymax=6,
				      enlargelimits,
				    ]
				    \addplot[thin,DarkRed,forget plot] table [x=snr, y=psgs, col sep=comma] {figs/m256_ray.csv};	
				\addplot[thin,DarkOrange,densely dotted] table [x=snr, y=cap, col sep=comma] {figs/m16_ray.csv};
				\addplot[thin,DarkBlue,densely dashed,forget plot] table [x=snr, y=qam, col sep=comma] {figs/m256_ray.csv};	
				    \end{axis}
				    \end{tikzpicture}%
				}}] at (pt) {};
							
			\end{tikzpicture}
	\caption{Mutual information achieved by the reference schemes \label{fig:mi_ray} and the learned joint probabilistic and geometric shaping on the Rayleigh channel. Magnification is done for $N = 256$.}
\end{figure}

A Rayleigh channel with \gls{LMMSE} estimation and equalization is considered in this section.
Fig.~\ref{fig:mi_ray} shows the lower bound on the capacity from~\cite[Corollary~1.3]{massivemimobook} (cf. orange dotted curve), and the mutual information achieved by the proposed joint shaping scheme (cf. red curve).
To the best of our knowledge, no optimal shaping scheme is known for the mismatched Rayleigh channel and, therefore, unshaped \gls{QAM} is considered as a baseline (cf. blue dashed plot).
One observes that the proposed joint-shaping approach enables significant gains over non-shaped \gls{QAM}.
An important aspect of this result is that no theoretical analysis or assumption on the channel was required to perform joint shaping.
This result is therefore encouraging to apply the proposed approach to other channels with untractable models.
Moreover, one could also perform shaping over non-differentiable or unknown channel models using recently proposed approaches~\cite{8792076} that remove the need of backpropagating gradients through the channel at training.

\section{Conclusion}
\label{sec:conclu}

This paper introduced a machine-learning based solution to the problem of constellation shaping.
In contrast to prior works on probabilistic shaping, the proposed scheme operates over a wide range of \glspl{SNR} and is not limited to specific families of distributions.
Instead, using the Gumbel-Softmax trick, a continuum of arbitrary distributions being function of the \gls{SNR} was learned.
As an extension, the presented approach was generalized to joint geometric and probabilistic shaping.
It was shown that joint shaping outperforms single geometric shaping or probabilistic shaping of \gls{QAM} regarding the achieved mutual information, and nearly reaches the capacity on an \gls{AWGN} channel.
On a Rayleigh channel, the proposed joint-shaping scheme outperforms unshaped \gls{QAM}.
The presented results are promising for other channel models, left as future research directions.

\appendix


The categorical cross entropy loss function can be rewritten as
\begin{align}
&\Lc(\thetav_S, \thetav_M,\thetav_D) \nonumber\\
&= -\int_{x} p_{\thetav_S,\thetav_M}(x) \int_{y} p(y|x) \log{\tilde{p}_{\thetav_D}(x|y)} dy dx  \nonumber\\
&= -\int_{x} \int_{y}  p_{\thetav_S,\thetav_M}(x,y)  \log{\tilde{p}_{\thetav_D}(x|y)} dy dx \nonumber\\
&= -\int_{x} p_{\thetav_S, \thetav_M}(x) \log{p_{\thetav_S,\thetav_M}(x)} dx \nonumber\\
& - \int_{x} \int_{y}  p_{\thetav_S,\thetav_M}(x,y)  \log{\frac{\tilde{p}_{\thetav_S,\thetav_M,\thetav_D}(x,y)}{p_{\thetav_S,\thetav_M}(y) p_{\thetav_S,\thetav_M}(x)}} dy dx \label{eq:l_ext}
\end{align}
where $p_{\thetav_S,\thetav_M}(x,y) = p_{\thetav_S,\thetav_M}(x)p(y|x)$, $p_{\thetav_S,\thetav_M}(y) = \int_{x} p_{\thetav_S,\thetav_M}(x,y)$, and $\tilde{p}_{\thetav_S,\thetav_M,\thetav_D}(x,y) = \tilde{p}_{\thetav_D}(x|y) p_{\thetav_S,\thetav_M}(y)$.

It is important to notice that $p_{\thetav_S,\thetav_M}(x,y)$ is the true joint distribution of $(X,Y)$, whereas $\tilde{p}_{\thetav_S,\thetav_M,\thetav_D}(x,y)$ is the joint distribution computed from the posterior approximated by the demodulator $\tilde{p}_{\thetav_D}(x|y)$.
Assuming that there is no $(s,t) \in \Sc^2$ such that $s \neq t$ and $f_{\thetav_M}(s) = f_{\thetav_M}(t)$, meaning that each symbol $s \in \Sc$ is uniquely mapped to a constellation point $x \in \CC$, then the first term of (\ref{eq:l_ext}) is the entropy of $S$, denoted by $H_{\thetav_S}(S)$.
One can see that
\begin{multline}
\Lc(\thetav_S, \thetav_M,\thetav_D) = 
H_{\thetav_S}(S) - I_{\thetav_S,\thetav_M}(X;Y)\\
+ \EE_{y} \LP \text{D}_{\text{KL}}\LB p_{\thetav_S,\thetav_M}(x|y)||\tilde{p}_{\thetav_D}(x|y) \RB \RP
\end{multline}
where D\textsubscript{KL} is the \gls{KL} divergence, and $p_{\thetav_S,\thetav_M}(x|y) = \frac{p_{\thetav_S,\thetav_M}(x,y)}{p_{\thetav_S,\thetav_M}(y)}$ is the true posterior of $X$.




\bibliographystyle{IEEEtran}
\bibliography{IEEEabrv,bibliography}

\begin{thebibliography}{10}
\providecommand{\url}[1]{#1}
\csname url@samestyle\endcsname
\providecommand{\newblock}{\relax}
\providecommand{\bibinfo}[2]{#2}
\providecommand{\BIBentrySTDinterwordspacing}{\spaceskip=0pt\relax}
\providecommand{\BIBentryALTinterwordstretchfactor}{4}
\providecommand{\BIBentryALTinterwordspacing}{\spaceskip=\fontdimen2\font plus
\BIBentryALTinterwordstretchfactor\fontdimen3\font minus
  \fontdimen4\font\relax}
\providecommand{\BIBforeignlanguage}[2]{{%
\expandafter\ifx\csname l@#1\endcsname\relax
\typeout{** WARNING: IEEEtran.bst: No hyphenation pattern has been}%
\typeout{** loaded for the language `#1'. Using the pattern for}%
\typeout{** the default language instead.}%
\else
\language=\csname l@#1\endcsname
\fi
#2}}
\providecommand{\BIBdecl}{\relax}
\BIBdecl

\bibitem{8054694}
T.~O’Shea and J.~Hoydis, ``An introduction to deep learning for the physical
  layer,'' \emph{{IEEE} Trans. on Cogn. Commun. Netw.}, vol.~3, no.~4, pp.
  563--575, Dec. 2017.

\bibitem{jones2018geometric}
R.~T. Jones, T.~A. Eriksson, M.~P. Yankov, B.~J. Puttnam, G.~Rademacher, R.~S.
  Luis, and D.~Zibar, ``{Geometric constellation shaping for fiber optic
  communication systems via end-to-end learning},'' \emph{arXiv preprint
  arXiv:1810.00774}, 2018.

\bibitem{jang2016categorical}
E.~Jang, S.~Gu, and B.~Poole, ``{Categorical Reparameterization with
  Gumbel-Softmax},'' \emph{Proc. Int. Conf. Learn. Represent. (ICLR)}, 2017.

\bibitem{7307154}
G.~{Böcherer}, F.~{Steiner}, and P.~{Schulte}, ``Bandwidth efficient and
  rate-matched low-density parity-check coded modulation,'' \emph{IEEE Trans.
  Commun.}, vol.~63, no.~12, pp. 4651--4665, Dec 2015.

\bibitem{kim2018communication}
H.~Kim, Y.~Jiang, R.~B. Rana, S.~Kannan, S.~Oh, and P.~Viswanath,
  ``Communication algorithms via deep learning,'' in \emph{Int. Zurich Seminar
  Inf. Commun. (IZS)}, Feb. 2018, pp. 48 -- 50.

\bibitem{kimOFDM2018}
M.~Kim, W.~Lee, and D.~H. Cho, ``A novel {PAPR} reduction scheme for {OFDM}
  system based on deep learning,'' \emph{{IEEE} Commun. Lett.}, vol.~22, no.~3,
  pp. 510--513, Mar. 2018.

\bibitem{osheamimo2017}
T.~J. O'Shea, T.~Erpek, and T.~C. Clancy, ``Physical layer deep learning of
  encodings for the {MIMO} fading channel,'' in \emph{Annu. Allerton Conf.
  Commun., Control, Comput. (Allerton)}, Oct. 2017, pp. 76--80.

\bibitem{bourtsoulatze2018deep}
E.~{Bourtsoulatze}, D.~B. {Kurka}, and D.~{Gündüz}, ``Deep joint
  source-channel coding for wireless image transmission,'' in \emph{IEEE Int.
  Conf. on Acoust., Speech and Signal Process. (ICASSP)}, May 2019, pp.
  4774--4778.

\bibitem{7322261}
P.~{Schulte} and G.~{Böcherer}, ``Constant composition distribution
  matching,'' \emph{IEEE Trans. Inf. Theory}, vol.~62, no.~1, pp. 430--434, Jan
  2016.

\bibitem{fritschek2019deep}
R.~Fritschek, R.~F. Schaefer, and G.~Wunder, ``Deep learning for channel coding
  via neural mutual information estimation,'' \emph{arXiv preprint
  arXiv:1903.02865}, 2019.

\bibitem{belghazi2018mine}
M.~I. Belghazi, A.~Baratin, S.~Rajeshwar, S.~Ozair, Y.~Bengio, A.~Courville,
  and D.~Hjelm, ``Mutual information neural estimation,'' in \emph{Proceedings
  of the 35th Int. Conf. on Mach. Learning}, vol.~80, 10--15 Jul 2018, pp.
  531--540.

\bibitem{8792076}
F.~{Ait Aoudia} and J.~{Hoydis}, ``Model-free training of end-to-end
  communication systems,'' \emph{IEEE J. Sel. Areas Commun.}, 2019.

\bibitem{hazan2012partition}
T.~Hazan and T.~Jaakkola, ``On the partition function and random maximum
  a-posteriori perturbations,'' in \emph{Proceedings of the 29th Int. Conf. on
  Mach. Learning}, 2012, pp. 1667--1674.

\bibitem{bengio2013estimating}
Y.~Bengio, N.~L{\'e}onard, and A.~Courville, ``{Estimating or Propagating
  Gradients Through Stochastic Neurons for Conditional Computation},''
  \emph{arXiv preprint arXiv:1308.3432}, 2013.

\bibitem{tensorflow2015-whitepaper}
\BIBentryALTinterwordspacing
``{TensorFlow}: Large-scale machine learning on heterogeneous systems,'' 2015,
  accessed: 2019-15-07. [Online]. Available: \url{http://tensorflow.org/}
\BIBentrySTDinterwordspacing

\bibitem{Kingma15}
D.~P. Kingma and J.~Ba, ``Adam: {A} method for stochastic optimization,'' in
  \emph{Proc. Int. Conf. Learn. Represent. (ICLR)}, May 2015, pp. 1--15.

\bibitem{kschischang1993optimal}
F.~R. Kschischang and S.~Pasupathy, ``Optimal nonuniform signaling for gaussian
  channels,'' \emph{IEEE Trans. Inf. Theory}, vol.~39, no.~3, pp. 913--929,
  1993.

\bibitem{massivemimobook}
\BIBentryALTinterwordspacing
E.~Bj\"{o}rnson, J.~Hoydis, and L.~Sanguinetti, ``Massive {MIMO} networks:
  {Spectral}, energy, and hardware efficiency,'' \emph{Foundations and
  Trends{\textregistered} in Signal Processing}, vol.~11, no. 3-4, pp.
  154--655, 2017. [Online]. Available:
  \url{http://dx.doi.org/10.1561/2000000093}
\BIBentrySTDinterwordspacing

\end{thebibliography}
\pagebreak
\end{document}